\documentclass[11pt,a4paper]{article}
\usepackage{amsmath, amsthm} 
\usepackage{amsfonts}
\usepackage{amssymb,graphics,psfrag}
\usepackage[bottom]{footmisc}
\usepackage{hyperref}
\usepackage{mathrsfs}
\usepackage{cite}
\usepackage{setspace}

\def\hybrid{\topmargin -10pt    \oddsidemargin 0pt
        \headheight 0pt \headsep 0pt
       \textwidth 6.25in       
      \textheight 9.5in       
        \marginparwidth .875in
        \parskip 5pt plus 1pt   \jot = 1.5ex}
\hybrid
\numberwithin{equation}{section}
\numberwithin{table}{section}
\begin{document}

\thispagestyle{empty}

\rightline{\small}

\vskip 3cm
\noindent
\begin{spacing}{1.5}
\noindent
{\LARGE \bf  Asymptotic symmetry groups and operator algebras }
 \end{spacing}
\vskip .8cm
\begin{center}
\linethickness{.06cm}
\line(1,0){447}
\end{center}
\vskip .8cm
\noindent
{\large \bf Waldemar Schulgin}

\vskip 0.2cm

{\em  \hskip -.05 cm Universit\'e Libre de Bruxelles and International  Solvay
Institutes}
\vskip -0.15cm
{\em \hskip -.05cm ULB-Campus Plaine CP231}
\vskip -0.15cm
{\em \hskip -.05cm B-1050 Brussels, Belgium}
\vskip -0.10cm
{\tt \hskip -.05cm waldemar.schulgin AT ulb.ac.be }
\vskip1cm

\noindent
{\large \bf Jan Troost}
\vskip 0.2cm
{\em \hskip -.05cm Laboratoire de Physique Th\'eorique\footnote{Unit\'e Mixte du CNRS et
    de l'Ecole Normale Sup\'erieure associ\'ee \`a l'universit\'e Pierre et
    Marie Curie 6, UMR
    8549.}}
    \vskip -.15cm
{\em \hskip -.05cm Ecole Normale Sup\'erieure}
 \vskip -.15cm
{\em \hskip -.05cm 24 rue Lhomond, 75005 Paris, France}

\vskip 1cm

\vskip0.6cm

\noindent {\sc Abstract:} We associate vertex operators to space-time
diffeomorphisms in flat space string theory, and compute their
algebra, which is a diffeomorphism algebra with higher derivative
corrections. As an application, we realize the
asymptotic symmetry group $BMS_3$ of three-dimensional flat space in terms of vertex
operators on the string worldsheet. This provides an embedding of the $BMS_3$
algebra in a consistent theory of quantum gravity.  Higher derivative corrections vanish
asymptotically. 
An appendix is dedicated to $\alpha'$ corrected
algebras in conformal field theory and string theory.

\newpage

{
\tableofcontents }

\newpage

\section{Introduction}
Holography in flat space has an ambiguous status. On the one hand,
there are arguments that quantum theories of gravity are holographic,
independently of the asymptotics of the space-time.  On the other
hand, concrete examples of quantum theories of gravity with a
holographic dual are confined to anti-de Sitter spaces, close
cousins, and low-dimensional examples. It thus remains worthwhile to
investigate to what extent holography in flat space holds. If it does,
one would like to construct explicit examples.

A guiding principle is the flat space asymptotic symmetry group. This
group first played a role in the analysis of gravitational waves in
flat 
space, and was later greatly elucidated
\cite{Bondi:1962px
,Sachs:1962wk
,Sachs:1962zza
,Penrose:1962ij
,Newman:1966ub
,Ashtekar:1996cd
,Barnich:2006av}.
For a review see e.g. \cite{Barnich:2010eb} and e.g. \cite{Bagchi:2012xr}
for further developments.  In particular, it was
shown that general relativity in three-dimensional flat space can be
given boundary conditions that 
allow for a large $BMS_3$ symmetry
algebra. This symmetry algebra is a contraction of two centrally
extended Virasoro algebras. As such, it is a close analogue to the
asymptotic symmetry group of $AdS_3$ \cite{Brown:1986nw}. The latter is suggestive of the
existence of a dual conformal field theory. It thus is natural to
analyze the $BMS_3$ algebra closely, with flat space holography in
mind.

In this paper, we want to analyze to what extent the $BMS_3$ symmetry
algebra can be embedded into a fully consistent theory of quantum
gravity.  We wish to show that tree level three-dimensional flat space
string theory represents the asymptotic symmetry group $BMS_3$. We
furthermore show that classes of possible $\alpha'$ corrections to the
vertex operator realization we propose are absent.

Our paper is organized as follows. In section \ref{ads3}, we 
recall how the $BMS_3$ algebra arises  from the asymptotic
symmetry algebra of $AdS_3$ in the limit of vanishing cosmological
constant (see e.g. \cite{Barnich:2012aw}). We discuss the differences between the case with zero and
negative cosmological constant, and discuss the subtleties in taking
the limit at the level of the vertex operator algebra on the
worldsheet.

In section \ref{algebra} we review how to associate vertex operators with
 diffeomorphisms in space-time, and we compute their algebra. It turns out that there are
possible $\alpha'$ corrections to the algebra. Since this is a perhaps surprising
feature of our analysis, we dedicate a long appendix to worldsheet vertex operator algebras,
both chiral and non-chiral, that exhibit $\alpha'$ corrections. We structured the
appendix such that it can be read independently.

We are further lead, in section \ref{algebra} to define the concept of
asymptotically marginal diffeomorphisms.  Indeed, we wish to remark
that all diffeomorphisms leave a given background on-shell.  In
string theory, each
diffeomorphism is associated to a BRST exact state, which therefore is
BRST closed and leaves the string field on-shell. We remind the reader
in appendix \ref{Q2} how this goes in practice in covariantly quantized
string theory (even when the diffeomorphisms
are not transversely polarized or marginal).  Still, such
diffeomorphisms have a non-trivial leg in the ghost sector. That
is why we will concentrate in subsection \ref{asmarg} on diffeomorphisms
that only shift the metric, and that are asymptotically transverse and
on-shell. They are associated to diffeomorphisms that can be written
as shifts of physical matter fields of the form $c \bar{c} O$ where
$O$ is a primary operator in the matter conformal field theory.

We then show in section
\ref{concasmarg} that the $BMS_3$ algebra can be represented by asymptotically marginal
vertex operators in string theory, and moreover, that potential
 $\alpha'$ corrections to the diffeomorphism algebra
vanish. We conclude in section \ref{concl} with a summary and a list of
topics for further analysis.

\section{The embedding in $AdS_3$}
\label{ads3}
In this section, we review how the asymptotic symmetry group of three-dimensional flat space arises from
that of anti-de Sitter space, and give motivation for constructing the $BMS_3$ algebra in flat space
string theory directly.
\subsection{The embedding in $AdS_3$ gravity}
A first approach to the problem of constructing the
$BMS_3$ algebra in string theory could consist in taking the limit of large radius of curvature
in the $AdS_3$ results. Indeed, this is possible, and leads from two copies of the Virasoro algebra
to the $BMS_3$ algebra, with central charge. We start out with two copies of the Virasoro
algebra with central charge $c$:
\begin{eqnarray}
{[} {\cal L}_m , {\cal L}_n {]} &=& (m-n) {\cal L}_{m+n} + \frac{c}{12} (m^3-m) \delta_{m+n,0}
\nonumber \\
{[} \tilde{{\cal L}}_m , \tilde {{\cal L}}_n {]} &=& (m-n) \tilde{{\cal L}}_{m+n} + \frac{c}{12} (m^3-m) \delta_{m+n,0}.
\end{eqnarray}
The classical general relativity value of the central charge is
$c=\frac{3}{2} \frac{l}{ G_N}$ where $l$ is the radius of curvature of the
$AdS_3$ space and $G_N$ is Newton's constant \cite{Brown:1986nw}. The
central charge acquires $\alpha'/l^2$ corrections in bosonic string theory
\cite{Troost:2010zz, Troost:2011ud}. We can obtain the $BMS_3$ algebra through the redefinition
\begin{eqnarray}
{\cal P}_m &=& \frac{1}{l} ( {\cal L}_m + \tilde{{\cal L}}_{-m})
\nonumber \\
{\cal J}_m &=&  {\cal L}_m - \tilde{{\cal L}}_{-m} \, ,
\end{eqnarray}
 and the contraction $l \rightarrow \infty$ with the charges ${\cal P}_m$ and ${\cal J}_m$ kept fixed, yielding:
\begin{eqnarray}
{[} {\cal J}_m , {\cal J}_n {]} &=& (m-n) {\cal J}_{m+n} 
\nonumber \\
{[} {\cal J}_m , {\cal P}_n {]} &=& (m-n) {\cal P}_{m+n} + \frac{c}{12} (m^3-m) \delta_{m+n,0}
\nonumber \\
{[} {\cal P}_m , {\cal P}_n {]} &=& 0,
\end{eqnarray}
where 
now $c = 3/G_N$. The limit
we performed can alternatively be described as the limit of large momentum compared
to the inverse radius of curvature of $AdS_3$.  
Note that 
the momentum charges ${\cal P}$ carry
dimension of one over length, as does the central charge.  
Without
reference scale (e.g. a momentum), the value of the central charge is
arbitrary. Combining momentum with the three-dimensional Newton
constant allows for the construction of a dimensionless ratio. We can
make this manifest in the algebra:
\begin{eqnarray}
{[} {\cal J}_m , {\cal J}_n {]} &=& (m-n) {\cal J}_{m+n} 
\nonumber \\
{[} {\cal J}_m , G_N {\cal P}_n {]} &=& (m-n)\,  G_N {\cal P}_{m+n} + \frac{1}{4} (m^3-m) \delta_{m+n,0}
\nonumber \\
{[} G_N {\cal P}_m , G_N {\cal P}_n {]} &=& 0.
\end{eqnarray}

\subsection{The embedding in $AdS_3$ string theory}
The worldsheet embedding of the $AdS_3$ general relativity asymptotic
symmetry algebra was performed in 
\cite{Giveon:1998ns, Kutasov:1999xu}. One can review
that calculation, and take the flat space limit at the very end to
recuperate the $BMS_3$ algebra, as we did above. That is one way to
embed the asymptotic symmetry algebra of flat space in string theory
-- by viewing flat space string theory as a limit of $AdS_3$ string
theory where we take the cosmological constant to zero.

We would like to develop a more direct route, working with strings in
flat space, and the flat space worldsheet string action. One reason is
the following. The calculation of \cite{Giveon:1998ns} corresponds to
an expansion in a free conformal field theory deformed by the operator
$\exp(- \rho/\sqrt{k})$, where the radial coordinate $\rho$ is taken
large and the radius of curvature squared over $\alpha'$ (i.e. the
level $k$) is kept fixed, such that one has a perturbative
expansion. Taking the zero cosmological constant limit $k \rightarrow
\infty$ does not commute with the large radius limit. The
non-commutativity of these two limits is one way to understand that
there is no direct way of adapting the calculation in
\cite{Giveon:1998ns} in $AdS_3$ to the flat space context
proper. Furthermore, we note that the $AdS_3$ conformal field theory
is interacting and difficult to solve while the flat space worldsheet
action is a free conformal field theory.  To learn about flat space
string theory and its asymptotic symmetry group, we want to develop
a more direct approach.

\section{The algebra of diffeomorphism vertex operators}
\label{algebra}

\subsection{Diffeomorphism vertex operators}

String theory is a theory of gravity, and is invariant under
diffeomorphisms.  The algebra of diffeomorphisms is part of the vast
symmetry algebra of string theory. In a covariant quantization,
diffeomorphisms correspond to shifts of the background by BRST exact worldsheet vertex
operators. In particular, diffeomorphisms correspond to graviton
vertex operators where the graviton fluctuation is taken to correspond
to the variation of the metric under an infinitesimal diffeomorphism,
at least to first order in $\alpha'$.  We will call such a vertex
operator a (generalized) diffeomorphism vertex operator. In the classical
general relativity limit of string theory, we expect the
diffeomorphism vertex operators to satisfy the diffeomorphism algebra. They
may satisfy an $\alpha'$ corrected algebra at higher order, but by abuse of nomenclature,
we will still call these operators diffeomorphism vertex operators.

More explicitly, there is a map from infinitesimal diffeomorphisms parameterized by a vector
field $\xi$ to graviton vertex operators $V_\xi$:
\begin{eqnarray}
V_\xi &=& \frac{1}{2 \pi \alpha'} \int d^2 z \, \delta G_{\mu \nu} \partial X^\mu \bar{\partial} X^\nu
\end{eqnarray}
where the variation of the metric $\delta G_{\mu \nu}$ is:
\begin{eqnarray}
\delta G_{\mu \nu} &=& \nabla_\mu \xi_\nu+\nabla_\nu \xi_\mu \, . \label{variation}
\end{eqnarray}
The first order commutator algebra of these diffeomorphism vertex operators is expected to 
reproduce the commutator algebra of diffeomorphisms, where the vector associated
to the commutator of diffeomorphisms $\xi_{1}$ and $\xi_2$ is the Lie bracket ${[} \xi_1, \xi_2 {]}^\mu=
\xi_1^\nu \partial_\nu \xi_2^\mu - \xi_2^\nu \partial_\nu \xi_1^\mu$. 

The structure of the algebra can be made more explicit
in the case of a flat space background. In $n$-dimensional flat space, the background
metric is trivial and the operator product expansions of the free coordinate fields $X^\mu$ are:
\begin{eqnarray}
X^\mu (z_1,\bar{z}_1) X^\nu (z_2,\bar{z}_2) & \approx & - \frac{\alpha'}{2} \log |z_1-z_2|^2 \, .
\end{eqnarray}
The covariant derivatives $\nabla_\mu$ in equation (\ref{variation}) reduce to ordinary
derivatives $\partial_\mu$.  We can use this information to calculate
the commutator of diffeomorphism vertex operators.
Before performing the calculation, we simplify the form of the
diffeomorphism vertex operators, using the worldsheet equations of motion and
Stokes theorem:
\begin{eqnarray}
V_\xi &=& \frac{1}{2 \pi \alpha'} \int d^2 z  \Big(\partial_\mu \xi_\nu+\partial_\nu \xi_\mu\Big) \partial X^\mu \, \bar{\partial} X^\nu
\nonumber \\
&=& \frac{1}{2 \pi \alpha'} \left( \oint dz \, \xi_\mu \partial X^\mu - \oint d \bar{z} \,  \xi_\mu \bar{\partial} X^\mu \right).
\end{eqnarray}
The contour integrals are performed over boundaries of the worldsheet, and around other insertions.
After these preliminaries, we are ready to compute the algebra of the (generalized)
diffeomorphism vertex operators.

\subsection{The commutator of diffeomorphism vertex operators}
In string theory it is natural to extend the algebra of diffeomorphisms by the algebra
of anti-symmetric gauge transformations. Indeed, T-duality symmetry of string theory puts them
on similar footing. We are thus motivated to define the following operators, reminiscent of
current components:
\begin{eqnarray}
j_z(\xi^L) &=& \xi^L \cdot \partial X
\nonumber \\
j_{\bar z}(\xi^R) &=& \xi^R \cdot \bar{\partial} X \, .
\end{eqnarray}
where $\xi^{L,R}$ are space-time fields. We start off with an important remark
on the algebra of these operators. The operators
$\xi^{L}$ and $\xi^{R}$ are functionals of the
coordinate fields $X^\mu$.  We can assume they can be Fourier
decomposed. The Fourier modes are exponentials with operator products:
\begin{eqnarray}
:e^{i k_1 \cdot X(z_1,\bar{z}_1)}: \ \,:e^{i k_2 \cdot X(z_2,\bar{z}_2)}: 
& = & |z_1-z_2|^{ \alpha' k_1 \cdot k_2} : e^{i k_1\cdot X  (z_1,\bar{z}_1) + i k_2\cdot X (z_2,\bar{z}_2)} : \, . 
\end{eqnarray}
The factor $|z_1-z_2|^{ \alpha' k_1 \cdot k_2}$ changes the dimension of the
product from the sum of the dimensions of the factor operators. The first important remark
we want to make is that almost always we will
ignore this type of contribution. That is restrictive, and we will discuss
the importance of this restriction on several occasions later on. In summary, in the following
we will often ignore $\xi$-$\xi$ contractions.

After this preliminary, we compute the algebra of diffeomorphisms and
anti-symmetric gauge transformations, through the operator product
expansion of the component operators.  There will be both double and
single contractions. The double contractions come with an extra power
of $\alpha'$ (which we set equal to two), and two extra
space-time
derivatives. The operator products are:
\begin{eqnarray}
j_z (z_1, \bar{z}_1)\,  j_z (z_2, \bar{z}_2) & \approx &
- \frac{1}{z_1-z_2} \xi^{L1 \rho} \, \partial_\rho \xi^{2L}_\mu \, \partial X^\mu (2)
+ \frac{1}{z_1-z_2} \xi^{2L \rho} \, \partial_\rho \xi^{1L}_\mu \,  \partial X^\mu (2)
\nonumber \\
& & 
- \frac{1}{(z_1-z_2)^2} \xi^{L1 \rho} (1)  \, \xi^{L2}_\rho (2)
- \frac{1}{(z_1-z_2)^2} \partial_\rho \xi^{L1 \nu} (1) \,  \partial_\nu \xi^{L2 \rho}  (2) 
 \nonumber 
\end{eqnarray}
where $(1)$ and $(2)$ denote operators evaluated at $z_1$ or $z_2$ respectively.
For the mixed operator product of current components we have:
\begin{eqnarray}
j_z (z_1, \bar{z}_1) \, j_{\bar{z}} (z_2, \bar{z}_2) & \approx &
- \frac{1}{z_1-z_2} \xi^{L1 \rho} \partial_\rho \xi^{2R}_\mu \, \bar{\partial} X^\mu(2)
+ \frac{1}{\bar{z}_1-\bar{z}_2} \xi^{2R \rho}\,  \partial_\rho \xi^{1L}_\mu  \partial X^\mu (2)
\nonumber \\
& & 
- {\rm{contact\ term}}
- \frac{1}{|z_1-z_2|^2} \partial_\rho \xi^{L1 \nu} (1) \,  \partial_\nu \xi^{R2 \rho}  (2) \, .
\nonumber 
\end{eqnarray}

\subsubsection*{The commutators}
We can use these operator products, valid at short distance on a
cylinder as well as on the plane to compute equal time commutators of
these operators. We evaluate products of operators at
$z=\sigma+i \tau$,
regularized by a split in the time direction $\tau$ in accord with time
ordering.\footnote{See e.g. section 5 of \cite{Ashok:2009xx} for the relevant techniques.}
After expanding arguments around
$\sigma_{1,2}$, we find:
\begin{eqnarray}
[ j_z (1) , j_z (2) ] &=& 2 \pi i \delta(\sigma_1-\sigma_2)
\Big(
- \xi^{L1 \rho} \partial_\rho \xi^{2L}_\mu \partial X^\mu
+  \xi^{2L \rho} \partial_\rho \xi^{1L}_\mu  \partial X^\mu
\Big) (\sigma_2)
\nonumber \\
& & + 2 \pi i \partial_{\sigma_1} \delta(\sigma_1-\sigma_2)
\Big(  \xi^{L1 \rho} (\sigma_1) \xi^{L2}_\rho (\sigma_2)
+  \partial_\rho \xi^{L1 \nu} (\sigma_1)  \partial_\nu \xi^{L2 \rho}  (\sigma_2)
\Big) 
 \end{eqnarray}
while  
the mixed commutator is:
\begin{eqnarray}
{[} j_z (1), j_{\bar{z}}(2) {]} & = &
2 \pi i \delta(\sigma_1 - \sigma_2)
\Big(
-  \xi^{L1 \rho} \partial_\rho \xi^{2R}_\mu\,  \bar{\partial} X^\mu
- \xi^{2R \rho} \partial_\rho \xi^{1L}_\mu \,  \partial X^\mu
\nonumber \\
& & 
 + \frac{1}{2} ( \partial_\mu \partial_\rho \xi^{L1 \nu}  \partial_\nu \xi^{R2 \rho}  
 \partial X^\mu
-  \partial_\mu \partial_\rho \xi^{L1 \nu}  \partial_\nu \xi^{R2 \rho} 
 \bar{\partial} X^\mu
\nonumber \\
& & 
- \partial_\rho \xi^{L1 \nu}  \partial_\mu \partial_\nu \xi^{R2 \rho}  
 \partial X^\mu
+   \partial_\rho \xi^{L1 \nu} \partial_\mu \partial_\nu \xi^{R2 \rho} 
 \bar{\partial} X^\mu
)
\Big) (\sigma_2) \, .
\end{eqnarray}
We now recall that diffeomorphisms and anti-symmetric gauge transformations
give rise to the vertex operators on the cylinder:
\begin{eqnarray}
D[\xi] &=& \frac{i}{4 \pi }  \int d \sigma \,  \Big( j_z (\xi) - j_{\bar z} (\xi)\Big) \, ,
\\
A[\tilde{\xi}] &=& \frac{i}{4 \pi  }  \int d \sigma\,  \Big( j_z (\tilde{\xi}) + j_{\bar z} (\tilde{\xi})\Big).
\end{eqnarray}
We obtain the commutators of the diffeomorphism charges after
double integration:
\begin{eqnarray}
{[} D[\xi_1] , D[\xi_2] {]}
&=& \frac{i}{4 \pi  } \int d \sigma \, 
\Big(
  \xi^{1 \rho} \partial_\rho \xi^{2}_\mu 
-   \xi^{2 \rho} \partial_\rho \xi^{1}_\mu  
\nonumber \\
&& + \frac{1}{2} \Big( \partial_\mu \partial_\rho \xi^{1 \nu}\,   \partial_\nu \xi^{2 \rho}  
- \partial_\rho \xi^{1 \nu}  \, \partial_\mu \partial_\nu \xi^{2 \rho}\Big)\Big)  
( \partial X^\mu
- \bar{\partial} X^\mu) \, .
\end{eqnarray}
At leading order, we find the expected diffeomorphism algebra.
At subleading order, we find a higher derivative
contribution. We can summarize the algebra of parameters:
\begin{eqnarray}\label{diffeoalgebra}
{[} \xi_1 , \xi_2 {]}_\mu  &=&   \xi^{1 \rho} \partial_\rho \xi^{2}_\mu
- \xi^{2 \rho} \partial_\rho \xi^{1}_\mu 
+ \frac{\alpha'}{4}  \Big(\partial_\mu \partial_\rho \xi^{1 \nu}\,   \partial_\nu \xi^{2 \rho}  
- \partial_\rho \xi^{1 \nu} \, \partial_\mu  \partial_\nu \xi^{2 \rho} \Big).
\end{eqnarray}
For the other brackets, we find:
\begin{eqnarray}
{[} A[\tilde{\xi}_1] , A[\tilde{\xi}_2] {]}&=& \frac{i}{4 \pi}  \frac{\alpha'}{4}\int d \sigma 
 \Big( -\partial_\mu  \partial_\rho \tilde{\xi}^{1 \nu} \,  \partial_\nu \tilde{\xi}^{2 \rho}  
+ \partial_\rho \tilde{\xi}^{1 \nu} \,  \partial_\mu  \partial_\nu \tilde{\xi}^{2 \rho}  
\Big)( \partial X^\mu
- \bar{\partial} X^\mu
)
\nonumber \\
{[} D[{\xi}_1] , A[\tilde{\xi}_2] {]} 
&=& \frac{i}{4 \pi} \int d \sigma 
\Big(
  \xi^{1 \rho} \partial_\rho \tilde{\xi}^{2}_\mu 
+  \partial_\mu \xi^{ 1 \rho}   \tilde{\xi}^{ 2}_\rho
+ \frac{\alpha'}{2} \partial_\mu \partial_\rho \xi^{1 \nu}\,  \partial_\nu \tilde{\xi}^{2 \rho}
\Big)  
(\partial X^\mu +
\bar{\partial} X^\mu
) 
\, .
\end{eqnarray}
We have found the algebra of pairs of parameters:
\begin{eqnarray}
{[}(\xi_1, \tilde{\xi}_1) , (\xi_2,\tilde{\xi}_2) {]}  &=& 
\Big([\xi_1 , \xi_2] ,  {\cal L}_{\xi_1} \tilde{\xi}_2 - {\cal L}_{\xi_2} \tilde{\xi}_1 
- c_0 \, 
d \big(\xi_1 \cdot \tilde{\xi}_2 -  \xi_2 \cdot \tilde{\xi}_1\big) \Big)\nonumber \\
& & 
+ \left(\frac{\alpha'}{4}  \Big(\partial_\mu \partial_\rho \xi^{1 \nu} \, \partial_\nu \xi^{2 \rho}  
- \partial_\rho \xi^{1 \nu} \,  \partial_\mu \partial_\nu \xi^{2 \rho} \Big)
- \frac{\alpha'}{4}  \Big(\partial_\mu \partial_\rho \tilde{\xi}^{1 \nu} \, \partial_\nu \tilde{\xi}^{2 \rho}  
- \partial_\rho \tilde{\xi}^{1 \nu} \, \partial_\mu \partial_\nu \tilde{\xi}^{2 \rho} \Big) ,\right.
\nonumber \\
& & 
\left. \frac{\alpha'}{2}  \partial_\mu \partial_\rho \xi^{1 \nu}\,  \partial_\nu \tilde{\xi}^{2 \rho}
-  \frac{\alpha'}{2}  
\partial_\mu \partial_\rho \xi^{2 \nu}\, \partial_\nu \tilde{\xi}^{1 \rho} 
 - c_1 \frac{\alpha'}{2} d \Big( \partial_\rho
\xi^{1 \nu} \partial_\nu \tilde{\xi}^{2 \rho} - \partial_\rho
\xi^{2 \nu} \partial_\nu \tilde{\xi}^{1 \rho}\Big)\right)\, .\nonumber\\
\end{eqnarray}
There are coefficients $c_0$ and $c_1$ which are total derivative
terms that are not fixed by our calculation.  At leading order, we
obtain the Courant bracket if we put $c_0 = 1/2$ as can be seen from
the first line. 
(This corresponds to a mid-point prescription for 
the right hand side of the commutator of currents \cite{Alekseev:2004np}.
It also provides the existence of an extra automorphism for the bracket \cite{Zwiebach:2011rg}.)
At higher order, the choice $c_1=1/2$ is equally
natural.  We fix these choices from here on.  At subleading order,
the algebra of our generalized diffeomorphism and anti-symmetric gauge
transformation vertex operators exhibits new features.  We find
$\alpha'$ corrections to the diffeomorphism algebra.  Generalized
anti-symmetric gauge transformations commute into a diffeomorphism
parameter. Et cetera. The algebra deserves further comment.

If we view diffeomorphisms and anti-symmetric gauge transformations as
given in terms of their standard definitions in terms of manifolds and
gerbes, there can be no corrections to their algebra. There is
moreover every indication that they form a symmetry group of string
theory at each order in the $\alpha'$ expansion of the effective
action. Here, we have mapped these classical symmetry generators to
quantum vertex operators in an old covariant approach, and we have found that the algebra of
operators at lowest order agrees with their geometric counterpart,
while at higher order, it receives corrections. 
There may exist charges (which are $\alpha'$ corrected) that will have an
uncorrected geometric algebra. Nevertheless, the operators we
define above are natural, and may form an alternative slice of the
large symmetry algebra of string theory that could also be useful. 
A proper embedding in a covariant BRST framework seems
primordial. These important points will turn out not to be crucial to the
particular application we have in mind in this paper. In spite of this, to
further frame the interesting questions that arise from these results, we give
a more extensive discussion of $\alpha'$ contributions to stringy
vertex operator algebras from the point of view of two-dimensional
chiral and non-chiral conformal field theory in the extended appendix
\ref{voa} to this paper. In this appendix, we review the chiral first order algebra of \cite{Losev:2005pu},
the chiral doubled algebra of \cite{Hohm:2013jaa}, and how they relate to the 
non-chiral algebra we determined above. Further analysis of our generic result may well be
fruitful.

\subsection{Other worldsheet quantum corrections}
\label{anom}
We return to remark on a point we left aside previously.  In our
calculation, we dropped an overall factor of $|z_1-z_2|^{\alpha' k_1
  \cdot k_2}$ in the operator product of Fourier modes with momenta
$k_1$ and $k_2$. To gauge the influence of this factor on the algebra
at lowest order,
we suppose that $\alpha' k_1 \cdot k_2 < < 1$. Then,
 we find
$|z_1-z_2|^{\alpha' k_1 \cdot k_2} \approx 1 + {\alpha' k_1 \cdot k_2}
\log |z_1-z_2|$.  Therefore in the commutator we will encounter
expressions of the form:
\begin{eqnarray}
\lim_{\epsilon \rightarrow 0} \frac{\log |\sigma-i \epsilon|}{\sigma-i \epsilon}
-\frac{\log |\sigma+i \epsilon|}{\sigma+i \epsilon}
& = & 2 \pi i \log \sigma \delta(\sigma) \, .
\end{eqnarray}
Still other contributions will be proportional to $\log \sigma
\delta'(\sigma)$.  These terms are of a different type than those that give rise
to the $\alpha'$ corrections we studied above, such that it is justified
to consider them separately. 
\subsection{Quantum corrections and marginality}
As reviewed in appendix \ref{Q2}, all diffeomorphism operators in
string theory, defined as the BRST operator acting on seed operators,
are manifestly BRST closed. However, if we want a diffeomorphism
operator to correspond strictly to a shift of a physical graviton
vertex operator (proportional to the left and right ghosts $c
\bar{c}$), then we must demand that the diffeomorphism is marginal,
namely, its momentum $k$ is on-shell ($k^2=0$) and transverse ($e
\cdot k=0$ where $e$ is the polarization of the vertex
operator). Let's recall how these conditions come about.  For
simplicity, we consider the left-moving part of a diffeomorphism
charge:
\begin{eqnarray}
V_\xi 
&=& \frac{1}{4 \pi} \oint dz \,\xi_\nu \partial X^\nu,
\end{eqnarray}
or more carefully, let us consider a Fourier component:
\begin{eqnarray}
V_{e,k} 
&=&  e_\nu (k) \, \partial X^\nu e^{i k X} \, .
\end{eqnarray}
We will compute when this operator is a primary of dimension $1$ on the left.
To that end, we compute the operator product expansion with the
left energy-momentum tensor of the theory. 
\begin{eqnarray}
-\frac{1}{2} \partial X^\mu \partial X_\mu (1) 
\cdot V_{e,k} (2)
&=& -i \frac{ e \cdot k}{(z-w)^3} e^{i k X} (2)
+ \frac{ \frac{\alpha' k^2}{4} + 1}{(z-w)^2} V_{e,k} (2)
+ \frac{1}{z-w} \partial V_{e,k}(2) \nonumber\\
\end{eqnarray}
which says that the operator $V_{e,k}$ is dimension one on the left on
the condition that the polarization is transverse ($e \cdot k=0$), and
the operator is on-shell ($k^2=0$).

If we consider the commutator of marginal diffeomorphism vertex
operators, and moreover require that the result also be a marginal
operator, then no anomalous worldsheet quantum corrections (of the
type discussed in subsection \ref{anom}) will occur.  Indeed, consider
two marginal diffeomorphisms, and require that they commute into a
third, marginal diffeomorphism. Marginality implies that $k_1$ and
$k_2$ square to zero. On the other hand, the newly generated
diffeomorphism has momentum $k_1+k_2$ and will only be marginal if
$(k_1+k_2)^2=0 = 2 k_1 \cdot k_2$.  The terms generated by
contractions of the parameters of diffeomorphism among themselves
will then vanish.  Note that if one starts with two marginal
diffeomorphisms, the result is not necessarily marginal, such that the
set of marginal diffeomorphism vertex operators does not
necessarily close among itself under commutation.

\subsection{Asymptotically marginal diffeomorphisms}
\label{asmarg}
For our purposes however, marginality will be too strong a requirement
on the diffeomorphism
vertex operators we wish
to consider.  Rather, we will consider diffeomorphisms that we call
{\em asymptotically marginal}. These are
 diffeomorphism operators that are
marginal on the worldsheet, only up to terms which are asymptotically
vanishing in space-time. This allows for sufficient freedom to realize
asymptotic symmetry groups in terms of worldsheet vertex operators.
These operators will act on the solution space with given asymptotic
boundary conditions. We illustrate these concepts with an example
in the next section.

\section{Three-dimensional flat space}
\label{flat} 
In this section, as an example of the use of the above concepts and results,
we study string theory in three-dimensional flat space and its
asymptotic symmetry algebra $BMS_3$ \cite{Barnich:2010eb}. The framework developed earlier
is valid for higher dimensional flat spaces as well.
We note that the $BMS_3$ diffeomorphisms generate the solution space in pure 
three-dimensional flat space gravity (as they do
in $AdS_3$ pure gravity) \cite{Barnich:2010eb}. Thus, these should correspond to 
asymptotically marginal diffeomorphisms. We will show that they do, and will show that
the classes of $\alpha'$ corrections computed previously are absent for these diffeomorphism
vertex operators. In this section, we suppose that string theory is compactified on an
appropriate space and that the non-compact directions correspond to the space-time
$\mathbb{R}^{2,1}$.

The worldsheet theory for a string in (Euclidean) three-dimensional flat space is:
\begin{eqnarray}
S &=& \frac{1}{2 \pi \alpha'} \int d^2 z \, \partial X^i \bar{\partial} X_i \, ,
\end{eqnarray}
where $i=1,2,3$.
 We will work in terms of the linearly related coordinate system of free
fields $(\phi,\gamma,\bar{\gamma})$:
\begin{eqnarray}
X^1 &=& \phi
\nonumber \\
X^2 + i X^3 &=& 
\gamma
\nonumber \\
X^2 - i X^3 &=& 
\bar{\gamma} \, .
\end{eqnarray}

\label{concasmarg}
\subsection{The $BMS_3$ diffeomorphism vertex operators}
We demand that the space-time be asymptotically Minkowski (see
e.g. \cite{Barnich:2010eb}).  Coordinates well adapted to the
calculation of the asymptotic symmetry group are $(u,r,\varphi)$ with
metric:
\begin{eqnarray}
ds^2&=&-du^2-2 du dr+r^2d\varphi^2 
\, .
\end{eqnarray}
In these coordinates, the boundary is at fixed $\varphi, u$ and radial
infinity $r\rightarrow\infty$. The string worldsheet theory is
interacting, and strongly interacting at large radius. A trick could
consist in introducing an auxiliary coordinate to reduce the $r^2$
term to a $r^{-2}$ interaction, but we will not follow this route
here.  Rather, we will go back and forth between the asymptotic symmetry
coordinates $(u,r,\varphi)$ and the free field coordinates
$(\phi,\gamma,\bar{\gamma})$. The metric in the latter coordinates is
\begin{equation}
ds^2=d\phi^2+d\gamma\, d\bar\gamma \ ,
\end{equation}  and the mapping between the coordinate sets is
\begin{eqnarray}
\phi&=&-i(u+r)\nonumber\\
\gamma&=& 
r e^{i\varphi}\nonumber\\
\bar\gamma&=& 
r e^{-i\varphi} \, .
\end{eqnarray}
The components of the vector fields representing the infinitesimal
diffeomorphisms that generate the $BMS_3$ algebra are \cite{Barnich:2010eb} :
\begin{eqnarray}\label{Killing}
\xi^u&=&T+uY'\nonumber\\
\xi^{\varphi}&=&Y-\frac{1}{r}\left(T'+uY''\right)\nonumber\\
\xi^r&=&-r Y'+T''+u Y''' \ ,
\end{eqnarray}
where the functions
$T$ and $Y$ are functions of the angular coordinate $\varphi$ only.
Primes on the functions
$Y$ and $T$ will correspond to the derivatives with respect to the $\varphi$ coordinate.

We wish to write down the vertex operators corresponding to the $BMS_3$ diffeomorphisms in the 
free field theory variables. To do this, we perform the coordinate transformation to the 
$(\phi,\gamma,\bar{\gamma})$ variables on the vector fields. Next, we Fourier decompose the
generators $T$ and $Y$ in $\varphi$. With those results, we can compute the corresponding
vertex operators, in free field variables. The results are as follows.

\subsubsection{The translation vertex operators}
We find the components of 
the translation diffeomorphisms corresponding to the $n$-th Fourier mode of $T$:
\begin{eqnarray}
\xi^{\phi}_{T,n}&=&-i(1-n^2)\left(\frac{\gamma}{\bar\gamma}\right)^{\frac{n}{2}}\, ,\nonumber\\
\xi^{\gamma}_{T,n}&=&-n(n-1)\left(\frac{\gamma}{\bar\gamma}\right)^{\frac{n+1}{2}}\, ,\nonumber\\
\xi^{\bar\gamma}_{T,n}&=&-n(n+1)\left(\frac{\gamma}{\bar\gamma}\right)^{\frac{n-1}{2}} \, .
\end{eqnarray}
The corresponding vertex operators $V_{T,n}$ are: 
\begin{eqnarray}
V_{T,n} &=& 
 \frac{1}{8 \pi} \oint dz
 \Bigg(  -2 i (1-n^2) \left(\frac{\gamma}{\bar\gamma}\right)^{\frac{n}{2}} \ {\partial} \phi
  - n (n+1) \left(\frac{\gamma}{\bar\gamma}\right)^{\frac{n-1}{2}} 
 \partial \gamma
  -   n (n-1) \left(\frac{\gamma}{\bar\gamma}\right)^{\frac{n+1}{2}} 
 \partial\bar\gamma\Bigg) 
 \nonumber \\
 &&
  -\frac{1}{8 \pi} \oint  d \bar z
 \Bigg(  -2i (1-n^2) \left(\frac{\gamma}{\bar\gamma}\right)^{\frac{n}{2}} \ {\bar\partial} \phi
  - n (n+1) \left(\frac{\gamma}{\bar\gamma}\right)^{\frac{n-1}{2}} 
 \bar\partial \gamma
  -   n (n-1) \left(\frac{\gamma}{\bar\gamma}\right)^{\frac{n+1}{2}} 
 \bar\partial\bar\gamma
 \Bigg) .
 \nonumber 
\end{eqnarray}

\subsubsection{The rotation vertex operators}
For the Fourier modes of infinitesimal rotation diffeomorphisms, we find:
\begin{eqnarray}\label{rotational}
\xi^{\phi}_{Y,n}&=&n \Big(
 i(1-n^2)  \phi+ (n^2-2) \sqrt{\gamma  \bar\gamma }\Big)\left(\frac{\gamma }{\bar\gamma}\right)^{\frac{n}{2}} \nonumber\\
 \xi^{\gamma}_{Y,n}&=& (n-1) \Big(n^2\phi+ i (n^2-1)\sqrt{\gamma 
   \bar\gamma }\Big)\left(\frac{\gamma }{\bar\gamma}\right)^{\frac{n+1}{2}} \nonumber\\
 \xi^{\bar\gamma}_{Y,n}&=&   (n+1)  \Big(n^2\phi+ i (n^2-1)\sqrt{\gamma 
   \bar\gamma }\Big)  \left(\frac{\gamma }{\bar\gamma}\right)^{ \frac{n-1}{2}} .
\end{eqnarray}
These components lead to the vertex operators $V_{Y,n}$ :
\begin{eqnarray}
V_{Y,n} &=& 
 \frac{1}{8 \pi} \oint dz
 \Bigg( 2 n
 \Big(
 i(1-n^2)  \phi+ (n^2-2) \sqrt{\gamma  \bar\gamma }\Big)\left(\frac{\gamma }{\bar\gamma}\right)^{\frac{n}{2}}
  \partial\phi\nonumber\\
&&
\qquad \qquad+ (n+1)  \Big(n^2\phi+ i (n^2-1)\sqrt{\gamma 
   \bar\gamma }\Big)  \left(\frac{\gamma }{\bar\gamma}\right)^{\frac{n-1}{2}}
 \partial \gamma\nonumber\\
&&
\qquad \qquad+ (n-1) \Big(n^2\phi+ i (n^2-1)\sqrt{\gamma 
   \bar\gamma }\Big)\left(\frac{\gamma }{\bar\gamma}\right)^{\frac{n+1}{2}}
 \partial \bar\gamma
 \Bigg )\nonumber\\
&&- \frac{1}{8 \pi} \oint d \bar z
 \Bigg( 2 n \Big(
 i(1-n^2)  \phi+ (n^2-2) \sqrt{\gamma  \bar\gamma }\Big)\left(\frac{\gamma }{\bar\gamma}\right)^{\frac{n}{2}}
  \bar\partial\phi\nonumber\\
&&
\qquad \qquad+ (n+1)  \Big(n^2\phi+ i (n^2-1)\sqrt{\gamma 
   \bar\gamma }\Big)  \left(\frac{\gamma }{\bar\gamma}\right)^{\frac{n-1}{2}}
  \bar \partial \gamma
\nonumber\\
 &&
\qquad \qquad+ (n-1) \Big(n^2\phi+ i (n^2-1)\sqrt{\gamma 
   \bar\gamma }\Big)\left(\frac{\gamma }{\bar\gamma}\right)^{\frac{n+1}{2}}
  \bar \partial \bar\gamma \Bigg )
 \, . 
\end{eqnarray}

\subsubsection{Asymptotic marginality}
\label{asmarg2}
We now wish to check whether the $BMS_3$ diffeomorphism vertex
operators satisfy the conditions of asymptotic marginality. We perform
this calculation in the free field flat space variables
$(\phi,\gamma,\bar{\gamma})$ in which we derived the conditions of transversality:
\begin{eqnarray}
\partial_\mu \xi^\mu & = & 0 
\end{eqnarray}
and
masslessness:
\begin{eqnarray}
\partial_\mu \partial^\mu \xi^\nu &=& 0 \, .
\end{eqnarray}
After a calculation, we find that transversality is satisfied exactly
for the translation as well as the rotation charges. The calculation
involves non-trivial cancellations between the coefficients, dependent
on the Fourier momentum $n$.  We moreover have that the massless, or
marginality condition is satisfied for the translation charges up to
terms that go like $r^{-2}$ and for the rotation charges up to terms
that go like $r^{-1}$. This comes about because the non-trivial
contributions arise from $\gamma$ and $\bar{\gamma}$ derivatives,
which lower the power of the radial coordinate $r$ in the
diffeomorphism parameter by $2$. Taking into account the leading term
in the parameter (which is $O(r^0)$ for translations and $O(r)$ for
rotations), we find the quoted suppression factors. {From} appendix
\ref{Q2} it should be clear that we are comparing the momentum squared
contribution in the worldsheet conformal dimension
to the leading contribution of $1$ arising from a worldsheet
derivative in the diffeomorphism vertex operator. Hence, at large
radius (compared to the string length), these terms indeed are
negligible.  Thus, the $BMS_3$ diffeomorphisms are asymptotically
marginal.

\subsection{The $BMS_3$ algebra}
\label{concalg}
We have already established that to leading order in $\alpha'$ the
(generalized, asymptotically marginal) diffeomorphism vertex operators that we
constructed satisfy the algebra of ordinary diffeomorphisms. Thus, for
the $BMS_3$ diffeomorphisms, they satisfy the $BMS_3$ algebra. We have
proven this by going to a coordinate system where the worldsheet
fields are free, then performing the operator products, and
commutators, after which we return to the coordinate system handy in
the definition and analysis of the asymptotics. Thus, the embedding of
the $BMS_3$ algebra in a consistent theory of quantum gravity has been
obtained. In this subsection, we show that the potential $\alpha'$
corrections we computed are subleading asymptotically, and that
therefore the $BMS_3$ algebra is represented without $\alpha'$ corrections.

To establish these facts, we first distinguish three types of commutators,
namely between translations, between rotations and translations, and
between rotations.
Firstly, as an illustrative example, we present the computation of the
higher derivative corrections to the algebra of rotational diffeomorphisms.  We
recall the algebra we derived in \eqref{diffeoalgebra}. We must be careful
to apply the formula to diffeomorphisms expressed in the free
field coordinates ($\phi,\gamma,\bar\gamma$).
Afterwards, we express the coefficients in the coordinate system $(r,u,\varphi)$
to simplify the task of comparing orders of coefficients in the limit of fixed $u, \varphi$ and $r\rightarrow
\infty$. In practice, we find:
\begin{eqnarray}
{[}\xi_{Y,n},\xi_{Y,m}]^\phi&=&-i(n-m)\xi_{Y,n+m}^{\phi}+\frac{im^2n^2(n^2-m^2)u^2 e^{i(m+n)\varphi}}{r}\nonumber\\&& +\frac{\alpha'}{4}\, \frac{i(n^2-m^2)(n^2+m^2-1)e^{i(n+m)\varphi}}{r}+{\cal{O}}\left(\frac{1}{r^2}\right)\, ,\nonumber\\
{[}\xi_{Y,n},\xi_{Y,m}]^{\gamma}&=&-i(n-m)\xi_{Y,n+m}^{\gamma}-\frac{m^2n^2(n^2-m^2)u^2 e^{i(m+n+1)\varphi}}{r}\nonumber\\&& +\frac{\alpha'}{4}\, \frac{(m-n)(m+n-1)(m^2+n^2-m-n-2)e^{i(n+m-1)\varphi}}{2r}+{\cal{O}}\left(\frac{1}{r^2}\right)\, ,\nonumber\\
{[}\xi_{Y,n},\xi_{Y,m}]^{\bar\gamma}&=&-i(n-m)\xi_{Y,n+m}^{\bar\gamma}-\frac{m^2n^2(n^2-m^2)u^2 e^{i(m+n-1)\varphi}}{r}\nonumber\\&& +\frac{\alpha'}{4}\, \frac{(m-n)(m+n+1)(m^2+n^2+m+n-2)e^{i(n+m+1)\varphi}}{2r}+{\cal{O}}\left(\frac{1}{r^2}\right)\, .\nonumber\\
\label{corrections}
\end{eqnarray}
All corrections are subleading with respect to the asymptotic algebra
of diffeomorphisms.
Indeed, comparing to the $\phi,\gamma,\bar\gamma$-components of the
rotational diffeomorphisms expressed in $r,u,\varphi$-coordinates:
\begin{eqnarray}
\xi_{Y,n}^\phi&=&-n\big(r+(n^2-1)u\big)e^{in\varphi}\nonumber\\
\xi_{Y,n}^\gamma&=&-i(n-1)\big(r+n^2 u\big)e^{i(n+1)\varphi}\nonumber\\
\xi_{Y,n}^{\bar\gamma}&=&-i(n+1)\big(r+n^2 u\big)e^{i(n-1)\varphi} \, , 
\end{eqnarray}
we see that the corrections in equation (\ref{corrections}) 
are subleading in the asymptotic region at fixed Fourier momentum.
Also, we find that the $\alpha'$ corrections of the form
$\partial_\mu \, \partial_\rho \xi^\nu\, \partial_\nu \xi^\rho$ are entirely
absent for translations, while for rotations acting on translations,
we need to compare the corrections to the translation diffeomorphisms:
\begin{eqnarray}
\xi^{\phi}_{T,n}&=&-i(1-n^2) e^{in\varphi}\, ,\nonumber\\
\xi^{\gamma}_{T,n}&=&-n(n-1)e^{i(n+1)\varphi}\, ,\nonumber\\
\xi^{\bar\gamma}_{T,n}&=&-n(n+1)e^{i(n-1)\varphi}
\, .
\end{eqnarray}
We find that the $\alpha'$ corrections are down by a power of $r$ with respect to these coefficients. 
We thus find that the asymptotic symmetry algebra is well represented by  our asymptotically marginal
vertex operators, even when including $\alpha'$ corrections.

Moreover, we note that the
condition of asymptotic marginality satisfied by $\xi_1$, $\xi_2$ and
their commutator, implies that the corrections proportional to $\alpha'
k_1 \cdot k_2$ are also absent. Alternatively, this follows from 
an argument
similar to the one 
given in subsection \ref{asmarg2} that guaranteed asymptotic marginality.
\subsection{The central charge}
Let's consider in more detail commutators that give rise to a central charge
contribution. These are commutators of the form:
\begin{eqnarray}
[ {\cal J}_m , {\cal P}_{-m} ] &=& 
2m { \cal P}_0 + \frac{c}{12} (m^3-m).
\end{eqnarray}
Thus, to see the central charge, we need to match up Fourier momenta. When Fourier
momenta match, we expect
terms in the commutator which are formally total derivatives.
However, we may generate operators like 
 $\partial_\sigma \varphi$, which can
give non-zero
contributions to the integral for a worldsheet profile with winding number $w$
around the $\varphi$ circle. The central charge will jump
when one crosses such a domain wall macroscopic string. This bears similarities to what
happens in $AdS_3$ \cite{Giveon:1998ns}. Let's see how this manifests in a string
theory context.

In $AdS_3$ solutions to string theory that arise from near-brane
limits of the F1-NS5 system, the central charge is given by $c=6 N_1
N_5$. Changing the number of fundamental strings $N_1$ by $\Delta N_1$
changes the central charge by $\Delta c = 6 \Delta N_1 N_5$
\cite{Giveon:1998ns}. We can think of our background as arising from
considering a large
number $N_5$  of NS5-branes, such
that the radius of curvature $R=\sqrt{N_5 \alpha'}$ becomes large with
respect to the string scale. One over the three-dimensional Newton constant,
meanwhile, is proportional to $N_1 \sqrt{N_5}$. It is fixed by an
attractor mechanism.  Changing the number of fundamental strings
considered in the near-brane limit corresponds then to a jump in the
three-dimensional Newton constant. It is this change in the central
charge that we detect in the worldsheet calculation.  The
normalization factor will be:
\begin{eqnarray}
\frac{1}{G_N} &=& 4 \frac{\sqrt{N_5} N_1}{\sqrt{\alpha'}} \, ,
\end{eqnarray}
such that the jump in the 
inverse Newton constant when we cross a macroscopic fundamental string
will be
$4 \sqrt{N_5} \Delta N_1 / \sqrt{\alpha'}$ where $\Delta N_1=w$ will correspond to the number
of times the macroscopic worldsheet wraps the angular direction $\varphi$. 

For the evaluation of the central charge in the $\mathbb{R}^{2,1}$
vacuum, a more elaborate analysis is necessary. It was performed in the
$AdS_3$ case in \cite{Troost:2011ud}.  The analogue of this calculation in $\mathbb{R}^{2,1}$,
as well as fleshing out our intuitive description of the change in central
charge when crossing a fundamental string, we leave for future work.

\section{Conclusion}
\label{concl}
We analyzed $\alpha'$ corrections to vertex operator algebras in
string theory, and to operator algebras in two-dimensional conformal
field theory. We applied our analysis of possible $\alpha'$
corrections to the $BMS_3$ algebra, the asymptotic symmetry algebra of
three-dimensional flat space. We showed that the algebra can be
represented in string theory, a consistent theory of quantum gravity, and
that potential higher derivative corrections are absent.  Many open research
directions have become more concrete. We name a few:
\begin{itemize}
\item Can one use an analogue of bulk-boundary propagators to construct an
exact version of the $BMS_3$ algebra, valid everywhere in the bulk 
(as in the case of $AdS_3$ \cite{Kutasov:1999xu}) ?
\item Give an interpretation of the representation theory of the $BMS_3$ algebra in terms
of the bulk gravitational theory.
\item Apply our formalism to other asymptotic symmetry algebras. The
  application to $BMS_4$, or four-dimensional Minkowski space is
  straightforward while curved bulk spaces require more work on worldsheet vertex operator
algebras.
\item Interpret the Jacobiators of $\alpha'$ corrected non-chiral vertex
operator algebras, and their space-time counterparts in an algebraic and in a 
geometric framework.
\item Embed our analysis in closed string field theory.
\item Analyze whether effective actions satisfy a constraint because of an $\alpha'$ 
corrected gauge algebra.
\item Extend/restrict our analysis to electromagnetism in three-dimensional flat space.
(See also \cite{StromingerStrings}.)
\item Extend the algebra to include supersymmetry.
\end{itemize}
Our work is but a step towards an improved understanding of how the
work on asymptotic symmetry groups in gravity is embedded in string
theory. One may legitimately hope that studying the symmetry of quantum theories of gravity
will further our understanding of holography.

\section*{Acknowledgments}
We would like to thank Costas Bachas, Glenn Barnich, Marc-Thierry Jaekel, 
Amir-Kian Kashani-Poor, Pierre-Henry Lambert and Giuseppe Policastro
for interesting discussions. 
The work of W.S. is partially supported  by a
Marina Solvay fellowship, by IISN - Belgium (conventions 4.4511.06
and 4.4514.08), by the ``Communaut\'e Fran\c{c}aise de Belgique"
through the ARC program and by the ERC through the ``SyDuGraM"
Advanced Grant. This work was supported in part by the ANR grant ANR-09-BLAN-0157-02.

\appendix

\section{Diffeomorphism operators are BRST exact}
\label{Q2}
In this appendix we discuss an elementary aspect of diffeomorphism vertex operators. Diffeomorphism
vertex operators arise from the action of the BRST operator $Q_B$ on a seed vertex operator $S$.
As such, they will be BRST closed, independent of the chosen seed $S$. In particular, there
will be no further constraint necessary in order for them to be on-shell in that sense.
 We wish to demonstrate this elementary fact explicitly.

We will concentrate on the left-movers only, and start out with a seed vertex operator $S$ which
is an exponential with momentum $k$:
\begin{eqnarray}
S &=& : e^{i k X} : \, ,
\end{eqnarray}
in a theory of free scalar fields $X$, i.e. flat space string theory.
Next, we compute the commutator with the BRST charge $Q_B$,
which is given as an integral over the BRST current $j_B$ (see e.g. \cite{Polchinski}):
\begin{eqnarray}
j_B &=& c T^{matter} + : b c \partial c : + \frac{3}{2} \partial^2 c
\nonumber \\
Q_B &=& \frac{1}{2 \pi i } \oint dz j_B.
\end{eqnarray}
The commutator gives the vertex operator $V$:
\begin{eqnarray}
V &=& [ Q_B, S(w)]
= \oint_{C_w} \frac{dz}{2 \pi i } j_B (z) S(w)
=c \partial S (w) + \frac{\alpha' k^2}{4} \partial c S(w).
\end{eqnarray}
as follows from the operator product expansion:
\begin{eqnarray}
T^{matter} (z) S(w) & \sim & \frac{ \frac{\alpha' k^2}{4}}{(z-w)^2} S(w) + \frac{1}{z-w} \partial S (w)
+ \dots
\end{eqnarray}
To confirm that this is a BRST exact vertex operator, we continue the analysis and
compute the commutator of the BRST operator $Q_B$ and the vertex operator $V$,
which is based on the operator product expansion of the energy momentum tensor with the
field $\partial S$ and the OPE of $b c \partial c$ with the ghost $c$ and its
derivative $\partial c$:
\begin{eqnarray}
T^{matter} (z) \partial S(w) & \sim & \frac{ \frac{\alpha' k^2}{2}}{(z-w)^3} S(w) + 
\frac{ \frac{\alpha' k^2}{4}+1}{(z-w)^2} \partial S(w)+
\frac{1}{z-w} \partial^2 S (w)
+ \dots
\nonumber \\
b\, c\,  \partial c (z)\,  c(w) & \sim & \frac{1}{z-w} c\,  \partial c (w) + \dots
\nonumber \\
b\,  c \, \partial c (z)\,  \partial c(w) & \sim & \frac{1}{(z-w)^2} c\,  \partial c (w)
+ \frac{1}{z-w} c\, \partial^2 c(w) + \dots 
\end{eqnarray}
The field $\partial S$ is a quasi-primary field. For the commutator we find:
\begin{eqnarray}
 [ Q_B, V(w)] &=& \partial^2 c\,  c \frac{\alpha' k^2}{4} S(w) + \partial c \, c
(\frac{\alpha' k^2}{4} +1) \partial S (w) + \frac{\alpha' k^2}{4} c\,  \partial c \, \partial S (w)
\nonumber \\
& & + c\,  \partial c \, \partial S (w) + \frac{\alpha' k^2}{4}\,  c\,  \partial^2 c \, S(w)
\nonumber \\
&=& 0.
\end{eqnarray}
{From} this exercise we see that there will be
no constraints on the momentum (or polarization) of a diffeomorphism vertex operator for
it to be BRST closed. In that sense, it is always on-shell.

We also see from (an easy extension of) 
the above calculation that if we wish the diffeomorphism vertex operator
to correspond to a shift only of the physical graviton vertex operator  proportional to
$c \bar{c} O$ where $O$ is a matter primary of dimensions $(1,1)$, then we must demand that the
seed operator $S$ is massless and  that the diffeomorphism
is transverse. Such a diffeomorphism vertex operator, we call marginal.

\section{Higher derivative corrections to operator algebras}
\label{voa}

In this appendix, we analyze algebras of currents and charges in
chiral and non-chiral conformal field theory, and their higher
derivative corrections.  Abstract algebras in
two-dimensional conformal field theories have been useful in finding
exact solutions to spectral problems as well as correlation
functions. They have very interesting connections with various
branches of mathematics, including affine Kac-Moody algebras,
generalized geometry and deformation theory.  They also serve as basic
building blocks for symmetries in string theory, including isometries
of target space, T-dualities, asymptotic symmetry groups as well as
gauge symmetries like diffeomorphisms. It is the latter application we
have in mind in the bulk of the paper.

In particular, we compute a number of these worldsheet algebras, and
their Jacobiator.  Firstly, we remark that the Jacobiator of chiral
current algebras is a total derivative by the theory of vertex
operator algebras \cite{Kac:1996wd}. Next, we compute the Jacobiator for a 
conformal field theory algebra with $\alpha'$ corrections associated
to a first order formalism for worldsheet sigma-models
\cite{Losev:2005pu}, and for an example based on the chiral
algebra of a free scalar field.\footnote{This algebra was
  independently calculated in \cite{Hohm:2013jaa} where the Jacobiator was
  computed as well. Moreover, \cite{Hohm:2013jaa} developed a theory
  of invariant tensors with applications to $\alpha'$ corrected
doubled geometry.} These
chiral algebras are warm-up examples for the non-chiral algebra we
study next, of diffeomorphism and anti-symmetric gauge transformation
operators.  In particular, we analyze non-chiral current algebras that
at leading order form a Courant algebra, and compute higher order corrections.
We also calculate the ensuing Jacobiator which contains total
derivative terms, and extra terms.

\subsection{Chiral algebras}
In this first subsection, we concentrate on chiral algebras, associated to (holomorphic) vertex
operator algebras.
When we analyze algebras of charges based on contour integrals of holomorphic
vertex operators, we can make good use of the mathematics of vertex operator
algebras (see e.g. \cite{Kac:1996wd} for a very readable
account). We can use the general theory to argue for the fact that
Jacobiators are total derivatives, and that the algebra of integrated charges
satisfies the Jacobi identity. In this section, we recall a few facts of the general theory of
vertex operators algebras and apply it to two algebras of charges. One example
is related to $\beta \gamma$ systems \cite{Losev:2005pu}, and another is based on the vertex operator
algebra of a chiral boson and was also discussed  in \cite{Hohm:2013jaa} recently.

\subsubsection{The general theory}
Vertex operator algebras contain a multitude of algebraic structures.
One is the $(n)$-product of vertex operators which to vertex operators
$a$ and $b$ associates the residue of the $(n+1)$st pole in the operator product of
mutually local vertex operators 
$a(z)$ and $b(w)$, expanded at $w$. That leads to the formula for mutually local
operators:
\begin{eqnarray}
a(z) b(w) &=& \sum_{j=0}^{N-1} \frac{a(w)_{(j)} b(w)}{(z-w)^{j+1}} + : a(z) b(w) :
\end{eqnarray}
We note that the zeroth product coincides with a contour integral action:
\begin{eqnarray}
a(w)_{(0)} b(w) &=& \oint_{w} dz \, a(z) b(w) \, ,
\end{eqnarray}
which in turn is equivalent to a commutator.
The associativity of the product of vertex operators evaluated at different
points implies a large number of properties of the $(n)$-products.
Using these properties, the Jacobiator of the 
$(0)$-product can be computed, and it is guaranteed to be a total derivative.
Explicitly, it is given by:
\begin{eqnarray}
a(w)_{(0)} \Big( b(w)_{(0)} c(w)\Big) + c_{(0)} \Big(a (w)_{(0)} b(w)\Big) +b(w)_{(0)} \Big( c(w)_{(0)} a(w)\Big)
&=& \nonumber \\
 - \sum_{j=1}^\infty (-1)^j \partial_w^{(j)} \bigg(c(w)_{(j)} \Big(a (w)_{(0)} b(w)\Big)
+ b(w)_{(0)}  \Big(c(w)_{(j)} a(w)\Big)\bigg). \label{jacobiatorvoa}
\end{eqnarray}
This is a standard result in the sense that it is known that modulo
the derivative of the vertex operator algebra, the Jacobi identity is
satisfied. Less well known seems to be the fact that the Jacobiator
can be calculated in terms of the $(n)$-products of the operators, and
that this identity is part of the structure of a strongly homotopy Lie
algebra \cite{Pinzon}. Strongly homotopy Lie algebras are known
to arise in covariant string field theory
\cite{Zwiebach:1992ie, Stasheff:1993ny}. It is gratifying to see
them feature in the elementary context of chiral conformal field theory as
well.

In the following, we give two examples of chiral vertex operator
algebras in which the zeroth product gives rise to an interesting
bracket operation on a set of vertex operators. Using the general
structure of vertex operator algebras, we will be guaranteed a total
derivative Jacobiator, which we compute. We are particularly interested
in bracket operations that contain higher derivative ($\alpha'$) corrections.

\subsubsection{An algebra in first order formalism}
\label{LMZ}
A first interesting chiral algebra that contains $\alpha'$ corrections
was discussed in \cite{Losev:2005pu}. It uses holomorphic vertex
operators only, and contour manipulations. 
We define two chiral vertex operators $p$ and $X$ which satisfy the following
operator products: 
\begin{eqnarray}
p_\mu(z) X^\nu(w) & \sim & - \frac{\delta^\nu_\mu}{z-w} \,.
\end{eqnarray}
We can think of the algebra as arising from a first order action principle for a free
chiral boson.  We now wish to analyze an algebra of charges and vertex operators
given by the following expressions:
\begin{eqnarray}
r (f,k) &=& \frac{1}{2 \pi i } \oint dz\,  f_\mu    e^{i k_\nu X^\nu} \partial X^\mu \, ,
\nonumber \\
n (e,k) &=& \frac{1}{2 \pi i} \oint dz \, e^\mu   e^{i k_\nu X^\nu} p_\mu  \, .
\end{eqnarray}
The relevant operator product expansion for the factors appearing in these charges are:
\begin{eqnarray}
p_\mu(z)  \,  e^{i k X (w)}  & \sim & - \frac{i k_\mu}{z-w}  \, e^{i k \cdot X(w)}  \, .
\end{eqnarray}
We will not only study the algebra of charges $r,n$ but also the vertex operator
algebra of their integrands, $R=  f_\mu \,   e^{i k_\nu X^\nu} \partial X^\mu$ and
$N= e^\mu  \, e^{i k_\nu X^\nu} p_\mu $.
In particular, let's consider the operator product of two operators $N$:
\begin{eqnarray}
N_1 (z_1) N_2(z_2)
 & \sim &
 \frac{1}{(z_1-z_2)^2}   e_1 \cdot  k_2\ e_2 \cdot k_1\, e^{i\left( k_1+k_2\right) \cdot X(z_2) } 
\nonumber \\
& & 
- \frac{i}{z_1-z_2}  \Big( e_1 \cdot k_2 \ e_2 \cdot p - e_2 \cdot k^1 \ e_1 \cdot p      \Big) e^{i (k_1+k_2)\cdot X} (z_2)  
\nonumber \\
& & 
 +  \frac{i}{z_1-z_2} e_1 \cdot  k_2\   e_2 \cdot k_1 \   k_1 \cdot  \partial X\   e^{i (k_1+k_2)\cdot X}  (z_2) 
\nonumber \\
& & 
+ \mbox{regular} 
\end{eqnarray} 
In the language of the $(n)$-products, we can reformulate 
the operator product expansion as:
\begin{eqnarray}
N_1 (z_2)_{(1)} N_2 (z_2) 
&=&    e_1 \cdot  k_2 \ e_2 \cdot k_1\  e^{i (k_1+k_2) X(z_2)}  
\nonumber \\
N_1 (z_2)_{(0)} N_2 (z_2) 
&=& - i \big( e_1 \cdot k_2 \ e_2 \cdot p - e_2 \cdot k_1\  e_1 \cdot p      \big) 
e^{i (k_1+k_2) X} (z_2) 
\nonumber \\
& & +   i e_1 \cdot  k_2 \ e_2 \cdot k_1  \ k_1 \cdot \partial X \  e^{i (k_1+k_2) X} (z_2) 
\nonumber 
\end{eqnarray}
We note that for holomorphic vertex operators, we have that the commutator of their
contour integrals is given by the contour integral of the $(0)$-product.
Thus, the properties of the $(0)$-product of integrands will largely determine
the properties of the charges (i.e. the integrated vertex operators).
The algebra of the vertex operators that appear as integrands in our charges is:
\begin{eqnarray}
R_{1(0)} R_2 = 0 \qquad
R_{(0)} N = i \, e \cdot k_1\  f \cdot \partial X\  e^{i (k_1+k_2) X} -  i\,  e \cdot f\  k_1 \cdot \partial X\ 
e^{i (k_1+k_2) X} \, . & &  
\end{eqnarray}
After Fourier transformation, we can define the operators $n$ 
and $r$ as \cite{Losev:2005pu}:
\begin{eqnarray}
n = \frac{1}{2 \pi i} \oint dz\,  v^\mu p_\mu & \qquad &
r = \frac{1}{2 \pi i} \oint dz\,  \omega_\mu \partial X^\mu \, ,
\end{eqnarray}
and find the non-zero commutators:
\begin{eqnarray}
[ n_{v_1}, n_{v_2} ] = n_{[ v_2 , v_1 ]} +  r_{\Omega (v_1,v_2)} \qquad
{[} r_\omega , n_v {]} = r_{{\cal L}_v \omega} & &  
\nonumber \\
\Omega_\mu (v_1,v_2) = -\frac{1}{2} \left(\partial_\mu\,  \partial_\nu v_1^\rho \, \partial_\rho v_2^\nu
- \partial_\mu \, \partial_\nu v_2^\rho \, \partial_\rho v_1^\nu\right), &&
\label{simple1}
\end{eqnarray}
where $\omega$ is considered a one-form and $v$ a vector.
Note that this agrees with the algebra of \cite{Losev:2005pu}.\footnote{Up to a minor typo in 
\cite{Losev:2005pu}, and a different convention for $\alpha'$.}
We used that we can neglect total derivatives in the parameter of the $r$-charge.
The algebra contains a higher
derivative $\alpha'$ correction. This is a basic example of the type of correction we wish to analyze.

\subsubsection*{The Jacobiator}
The only non-trivial Jacobiator is the one where we consider three
$n$-operators.  Using the $(j)$-product formalism, we see that the
Jacobiator contains terms with $(1)$-products at most.  We thus have
for the Jacobiator $ [n_1,[n_2,n_3]] + \mbox{cyclic} = r_\omega $
where $\omega$ is a total derivative determined by (see formula
\ref{jacobiatorvoa}):
\begin{eqnarray}
& &  \partial_w \bigg(
V_3(w)_{(1)} \Big(V_1 (w)_{(0)} V_2(w)\Big)
+
V_2 (w)_{(0)} \Big(V_3(w)_{(1)} V_1(w)\Big)\bigg)
\nonumber \\
&=& 
\partial_w \bigg( V_3 (w)_{(1)} \Big( - i  \big(e_1 \cdot k_2 \ e_2 \cdot p - e_2 \cdot k_1 \ e_1 \cdot p\big)
e^{i k_1+k_2) X} (w)\Big)
\nonumber \\
& & 
+ V_2(w)_{(0)} \Big( e_3 \cdot k_1\  e_1 \cdot k_3 \ e^{i (k_1+k_3) X(w)} \Big) \bigg) \nonumber \\
&=& \Big(- i\,  e_1 \cdot k_2\ e_2 \cdot k_3\ e_3 \cdot k_1
+ i \, e_1 \cdot k_3\  e_2 \cdot k_1 \ e_3 \cdot k_2
- i \, e_1 \cdot k_2\  e_2 \cdot k_3 \ e_3 \cdot k_2
\nonumber \\
& & 
- i \, e_1 \cdot k_3 \ e_2 \cdot k_3\  e_3 \cdot k_1-i \, e_1 \cdot k_2 \ e_2 \cdot k_1 \ e_3 \cdot k_1\Big)
\partial_w e^{i (k_1 + k_2 + k_3) X(w)} \, .
\end{eqnarray}
We conclude that in the vertex operator
formalism, the Jacobiator has total derivative parameter $\omega$ given by\footnote{Our convention for anti-symmetrization 
is $[ab]=ab-ba$.}:
\begin{eqnarray}
\omega_{\mu} &=& \partial_\mu S_{1st}
\nonumber \\
S_{1st} &=&
\frac{1}{3} \partial_\nu \xi_{[1}^\rho\,  \partial_\rho \xi_{2}^\sigma\,  \partial_\sigma \xi_{3]}^\nu
+ ( \xi_{1}^\rho \,  \partial_\rho\, \partial_\nu   \xi_2^\sigma\, \partial_\sigma \xi_{3}^\nu
+ \mbox{cycl. perm} )
\end{eqnarray}
It is crucial to remark that the total derivative Jacobiator depends on the choice
of total derivative terms in the commutator. To reproduce the above Jacobiator from the
brackets, one would use the choice of total derivative terms:
\begin{eqnarray}
[ n_{v_1}, n_{v_2} ] = n_{[ v_2 , v_1 ]} +  r_{\Omega (v_1,v_2)} \qquad
{[} n_v , r_\omega {]} = -r_{v^\rho \partial_\rho \omega_\mu +\partial_\mu v^\rho \omega_\rho }
\qquad
\Omega_\mu (v_1,v_2) = -\partial_\mu \partial_\nu v_1^\rho \partial_\rho v_2^\nu 
\, .
\nonumber 
\end{eqnarray}
This choice is dictated by the fact that the $\mu$ derivative, arising from the Taylor expansion
of an operator, by convention in vertex operator algebras, is always performed on the first
operator, and therefore acts on the first parameter in the commutator brackets.
We can simplify the Jacobiator by making a particular choice of brackets. Several
choices give a simple Jacobiator result. One choice is the original one we made in
equation (\ref{simple1}), while another choice is:
\begin{eqnarray}
[ n_{v_1}, n_{v_2} ] = n_{[ v_2 , v_1 ]} +  r_{\Omega (v_1,v_2)} \qquad
{[} n_v , r_\omega {]} = -r_{v^\rho \partial_\rho \omega_\mu -\partial_\mu  \omega_\rho v^\rho}
\qquad
\Omega_\mu (v_1,v_2) = -\partial_\mu \partial_\nu v_1^\rho \partial_\rho v_2^\nu 
\, .
\nonumber 
\end{eqnarray}
In both cases the seed function is proportional to:
\begin{eqnarray}
S_{1st} &=& \frac{1}{3} \partial_\nu \xi_{[1}^\rho \, \partial_\rho \xi_{2}^\sigma\,  \partial_\sigma \xi_{3]}^\nu
\end{eqnarray}
It has the distinguishing feature of being anti-symmetrized over the indices $1,2,3$.
The ambiguity in the Jacobiator will be present in all future computations. In the following,
we will prefer to work with brackets which are anti-symmetrized, and which correspond
(when relevant) to Courant brackets at lowest order.
It is tedious but straightforward to work out
the Jacobiators for all other choices of total derivative terms.

\subsubsection{A purely left gauge theory algebra}
In this subsection, we want to study a second example of a chiral
algebra of currents and charges.  The algebra is embedded in the vertex operator
algebra of a free chiral scalar field $X$. We can think of $X=X_L$ as
holomorphic, and we will again denote its left momentum $k_L$ by $k$ in this
section.  We  want to compute the (chiral) algebra of the
operators:
\begin{eqnarray}
Q_L &=& \frac{1}{2 \pi i} \oint \xi^L(X) \cdot \partial X \,.
\end{eqnarray}
However, now we work with the elementary operator product:
\begin{eqnarray}
X^\mu(z) X^\nu(w) & \sim & - \eta^{\mu \nu} \log (z-w) \, .
\end{eqnarray}
After Fourier decomposition, and noting the operator product equality
\begin{eqnarray}
: e^{i k_i \cdot X(z)} :\ : e^{i k_j \cdot X(w)} : & = & z^{\frac{\alpha'}{2} k_i \cdot k_j} : 
e^{ i k_i \cdot X(z)} e^{i k_j \cdot X(w)}: \, 
\end{eqnarray}
we see that to have a strict vertex operator algebra, we need $k_i \cdot k_j \in
2 \mathbb{Z}$. In the following, we make the stronger assumption
$k_i \cdot k_j =0$, and we will remark on it when pertinent.
Another difference with the algebra
of subsection \ref{LMZ} will be that additional terms are generated
because $\partial X$ will contract with $\partial X$ and $e^{i k X}$. After a calculation very similar
to that of the previous subsection, we find the commutator
of these charges:
\begin{eqnarray}
[ Q_L(\xi_1^L) , Q_L(\xi_2^L) ] &=& 
- Q_L ( [\xi_1^L,\xi_2^L] )
+ Q_L ( \omega (\xi_1^L,\xi_2^L))
\label{LL}
\end{eqnarray}
where we have defined:
\begin{eqnarray}
\omega_\mu (\xi_1^L, \xi_2^L)
&=& -\frac{1}{2}\Big( \partial_{\mu}   \xi_1^L\cdot \xi_2^L-\partial_{\mu}   \xi_2^L\cdot \xi_1^L\Big)
  +\frac{\lambda_3}{2} \Big(\partial_\mu \partial_\rho \xi_1^{L \sigma} \partial_\sigma \xi_2^{L \rho}-\partial_\mu \partial_\rho \xi_2^{L \sigma} \partial_\sigma \xi_1^{L \rho}\Big)  \, ,\nonumber\\
\end{eqnarray}
and $\lambda_3$ is proportional to $\alpha'$.
Again, the Jacobi identity will be satisfied by the general theory of vertex operator algebras. If we
record the algebra in terms of the parameters of the charges, 
we can compute the parameter of the Jacobiator
to be a total derivative. It is explicitly given by :
\begin{eqnarray}
J_{L} &=& \partial_\mu S_{L}
\nonumber \\
S_{L} &=&
\frac{1}{8}\, \Big(\xi^{[1\nu}\, \xi^{2\rho}\, K^{3]}_{\nu\rho}+\lambda_3\, \xi^{[1\nu}\, K^{2\rho\sigma}\partial_\nu K^{3]}_{\rho\sigma}+\frac{2\lambda_3}{3}\, {K^{[1}_\nu}^\rho \,  {K^2_\rho}^\sigma \,  {K^{3]}_\sigma}^\nu\Big) 
\end{eqnarray}
where 
\begin{equation}
K_{\mu\nu}=\partial_\mu\xi_{\nu}-\partial_\nu\xi_{\mu} \, .
\end{equation}
An identical right algebra associated to an anti-holomorphic vertex
operator algebra can be constructed.  In our final formula, we
put the Jacobiator in a form that is easy to match to
\cite{Hohm:2013jaa}.

\subsubsection*{Bootstrapping a purely left algebra}
In this subsection, we take an alternative approach to finding higher derivative corrections.
We analyze a class of extensions of the lowest order algebra and demand that the extension still
satisfy the Jacobi identity for charges (or that the Jacobiator of parameters is a total
derivative).
We set up the problem with the ansatz:
\begin{eqnarray}
[ Q_L(\xi_1^L) , Q_L(\xi_2^L) ] &=& 
- Q_L ( [\xi_1^L,\xi_2^L] )
+ Q_L ( \omega (\xi_1^L,\xi_2^L; \lambda_1,\lambda_3,\epsilon_1,\epsilon_3))
\end{eqnarray}
where we define:
\begin{eqnarray}
\omega_\mu (\xi_1^L, \xi_2^L; \lambda_1,\lambda_3,\epsilon_1,\epsilon_3)
&=& -\frac{\lambda_1}{2}\left( \partial_\mu \xi_1^L \cdot \xi_2^L -\partial_\mu \xi_2^L \cdot \xi_1^L \right)+\epsilon_1 \, \partial_\mu \left(\xi_1^L\cdot\xi_2^L\right)\nonumber\\&&-
\frac{\lambda_3}{2}\left( \partial_\mu\,  \partial_\rho \xi_1^{L \sigma} \, \partial_\sigma \xi_2^{L \rho} -\partial_\mu \, \partial_\rho \xi_2^{L \sigma}\,  \partial_\sigma \xi_1^{L \rho} \right)
+\epsilon_3\, \partial_\mu\left(\partial_\rho \xi_1^{L \sigma} \, \partial_\sigma \xi_2^{L \rho}\right)
\, .
\nonumber\\
\end{eqnarray}
We want to check the Jacobi identity for generic parameters
$\lambda_1$, $\lambda_3$ and the impact of the total derivatives which
we can add to $\omega_\mu$.
We can then prove the following results.  When $\lambda_3 = 0$, we
find that $\lambda_1=0$ or $\lambda_1=1$ are the only two solutions to
the Jacobi identity, independently of the values of
$\epsilon_1,\epsilon_3$.  If we have that $\lambda_3 \neq 0$, then we
must have $\lambda_1=1$. 
 In other words, we
cannot add the $\lambda_3$ term without adding the $\lambda_1=1$
term. On the dimensionful parameter $\lambda_3$ there is no further
condition. The anti-symmetric choice of total derivative term
($\epsilon_1=\epsilon_3=0$) matches the bracket in equation (\ref{LL}).

\subsection{A non-chiral algebra}
In this section, we analyze how the algebras are modified when the left- and right-moving
sectors communicate. The general theory of vertex operator algebras, and their tensor products,
will now no longer guarantee a total derivative Jacobiator. We can thus expect new algebraic
structures. We focus on the example of the algebra of vertex operators associated
to diffeomorphism and anti-symmetric gauge parameters.

First of all, we remark that a pedagogical classical derivation of
the Courant bracket of local invariances on the string worldsheet
is given in \cite{Alekseev:2004np}. Moreover, a clear explanation
of the Courant bracket and why it arises from non-commutativity of
diffeomorphisms and anti-symmetric
gauge transformations and that it has enhanced symmetry properties 
has been reviewed in \cite{Zwiebach:2011rg}, where further original references
can be found.

Here we will be interested in higher derivative corrections to the algebra
of (generalized) diffeomorphism vertex operators. To simplify our life, we
will consider a model of chiral bosons, both left and right, and ignore contact
terms. Thus, we consider the following algebras of chiral fields:
\begin{eqnarray}
X_L (z) X_L(w) & \sim & - \log (z-w)
\nonumber \\
X_R(\bar{z}) X_R(\bar{w}) & \sim & - \log (\bar{z} - \bar{w}) \, .
\end{eqnarray}
To mix the left and the right algebras in this model, we introduce the following charges:
\begin{eqnarray}
Q_L &=& \oint\,  dz \xi_L\Big(X_L(z), X_R(\bar{z})\Big) \partial X_L (z)
\nonumber \\
Q_R &=& \oint \, d \bar{z} \xi_R\Big(X_R(\bar{z}), X_L(z)\Big) \bar{\partial} X_R (\bar{z}) \, .
\end{eqnarray}
We leave the basic operator product expansions unchanged. 
The resulting algebra is:
\begin{eqnarray}
[ Q_L(\xi_1^L) , Q_L(\xi_2^L) ] &=& 
 Q_L ( [\xi_1^L,\xi_2^L] )
+ \frac{1}{2} Q_L \left( \partial_\mu^L \xi_1^\rho \xi_{2 \rho} - \partial_\mu^L \xi_2^\rho \xi_{1 \rho} \right)
- \frac{1}{2} Q_R  \left( \partial_\mu^R \xi_1^\rho \xi_{2 \rho} - \partial_\mu^R \xi_2^\rho \xi_{1 \rho} \right)
\nonumber \\
& & 
+ \frac{\alpha'}{4} Q_L \left( \partial_\mu^L \partial^L_\sigma \xi_1^\rho \, \partial^L_\rho \xi_{2}^\sigma 
                    -\partial_\mu^L \partial_\sigma^L \xi_2^\rho \, \partial_\rho^L \xi_{1}^\sigma \right)
- \frac{\alpha'}{4} Q_R  \left( \partial_\mu^R \partial_\sigma^L  \xi_1^\rho\,  \partial_\rho^L \xi_{2}^\sigma
                    -\partial_\mu^R \partial_\sigma^L \xi_2^\rho\,  \partial_\rho^L \xi_{1}^\sigma \right) \, ,
\nonumber
\end{eqnarray}
and an identical right algebra (with left and right interchanged).
The left-right algebra is
\begin{eqnarray}
[ Q_L(\xi_1^L) , Q_R(\xi_2^R) ] &=&
- Q_L \left( \xi^{R 2\rho} \partial_\rho^R \xi^{L1}\right)
+ Q_R \left( \xi^{L1  \rho} \partial_\rho^L \xi^{R2}\right)
\nonumber \\
& & 
+ \frac{\alpha'}{4} Q_L\left ( \partial_\mu^L \partial_\rho^R \xi^{L 1 \nu} \partial_\nu^L \xi^{R 2 \rho}
-  \partial_\mu^L \partial_\rho^L \xi^{R 2 \nu} \partial_\nu^R \xi^{L 1 \rho}
\right)\nonumber \\
& & 
+ \frac{\alpha'}{4} Q_R \left(  \partial_\mu^R \partial_\rho^R \xi^{L 1 \nu} \partial_\nu^L \xi^{R 2 \rho}
-  \partial_\mu^R \partial_\rho^L \xi^{R 2 \nu} \partial_\nu^R \xi^{L 1 \rho}
\right) \, .
\end{eqnarray}
This algebra is useful in doubled geometry applications.
When we identify derivatives with respect to left and right coordinates, 
we find the algebra:
\begin{eqnarray}
[ Q_L(\xi_1^L) , Q_L(\xi_2^L) ] &=& 
  Q_L ( [\xi_1^L,\xi_2^L] )
+ \frac{1}{2} Q_L ( \partial_\mu \xi_1^\rho \xi_{2 \rho} - \partial_\mu \xi_2^\rho \xi_{1 \rho} )
- \frac{1}{2} Q_R  ( \partial_\mu \xi_1^\rho \xi_{2 \rho} - \partial_\mu \xi_2^\rho \xi_{1 \rho} )
\nonumber \\
& & 
+ \frac{\alpha'}{4} Q_L ( \partial_\mu \partial_\sigma \xi_1^\rho \partial_\rho \xi_{2}^\sigma 
                    -\partial_\mu \partial_\sigma \xi_2^\rho \partial_\rho \xi_{1}^\sigma )
- \frac{\alpha'}{4} Q_R  ( \partial_\mu \partial_\sigma \xi_1^\rho \partial_\rho \xi_{2}^\sigma
                    -\partial_\mu \partial_\sigma \xi_2^\rho \partial_\rho \xi_{1}^\sigma ) \, .
\nonumber 
\end{eqnarray}
We also find the mixing: 
\begin{eqnarray}
[ Q_L (\xi_1^L) ,  Q_R(\xi_2^R) ] &=&  
- Q_L ( \xi^{R 2\rho} \partial_\rho \xi^{L1})
+ Q_R ( \xi^{L1  \rho} \partial_\rho \xi^{R2})
\nonumber \\
& & 
+ \frac{\alpha'}{4} Q_L ( \partial_\mu \partial_\rho \xi^{L 1 \nu} \partial_\nu \xi^{R 2 \rho}
-  \partial_\mu \partial_\rho \xi^{R 2 \nu} \partial_\nu \xi^{L 1 \rho}
)\nonumber \\
& & 
+ \frac{\alpha'}{4} Q_R (  \partial_\mu \partial_\rho \xi^{L 1 \nu} \partial_\nu \xi^{R 2 \rho}
-  \partial_\mu \partial_\rho \xi^{R 2 \nu} \partial_\nu \xi^{L 1 \rho}
) \, .
\end{eqnarray}
This is the $\alpha'$ corrected algebra of generalized anti-symmetric gauge transformation and diffeomorphism
vertex operators in a left-right separated basis.
To go to a standard basis, we use the linear map:
\begin{eqnarray}
D(\xi) =\frac{1}{2}\Big( Q_L(\xi) + Q_R(\xi)\Big) & \quad & A(\tilde{\xi}) =\frac{1}{2}\Big( Q_L(\tilde{\xi}) - Q_R(\tilde{\xi})\Big)
\nonumber \\
Q_L(\xi_L) = D(\xi_L)+A(\xi_L) & \quad & Q_R(\xi_R) = D(\xi_R) - A(\xi_R) \, ,
\end{eqnarray}
and find:
\begin{eqnarray}
{[} D(\xi^1) , D(\xi^2) ] &=&
D( [\xi^1,\xi^2]^\mu ) + \frac{1}{2} D \Big(\partial^\mu \, \partial_\rho \xi^{1 \sigma} \partial_\sigma \xi^{2 \rho}-\partial^\mu \,\partial_\rho \xi^{2 \sigma} \partial_\sigma \xi^{1 \rho}\Big)
\nonumber \\
 {[} A(\tilde{\xi}^1) , A(\tilde{\xi}^2) ] &=&
 -\frac{1}{2} D \Big(\partial^\mu\,  \partial_\rho \tilde{\xi}^{1 \sigma} \partial_\sigma \tilde{\xi}^{2 \rho}
-\partial^\mu\,  \partial_\rho \tilde{\xi}^{2 \sigma} \partial_\sigma \tilde{\xi}^{1 \rho}\Big)
\nonumber \\
{[} D(\xi^1) , A(\tilde{\xi}^2) ] &=&
A\Big( {\cal L}_{\xi^1} \tilde{\xi}^2-\frac{1}{2} d \,( {\xi^1} \cdot \tilde{\xi}^2)
+ \frac{1}{2}  \Big(\partial^\mu \, \partial_\rho \xi^{1 \sigma} \partial_\sigma \tilde{\xi}^{2 \rho}-\partial^\mu\,  \partial_\rho \tilde{\xi}^{2 \sigma} \partial_\sigma \xi^{1 \rho}\Big)\Big) \, .\nonumber\\
\end{eqnarray}
The Jacobiators in this basis are: 
\begin{eqnarray}
[D(\xi^1),[D(\xi^2),D(\xi^3)]]+{\rm cycl.\ perm.}&=& D\left[
- 
\frac{1}{6}\partial_\mu\Big(
\partial_\alpha\xi^{[1\beta}\,\partial_\beta\xi^{2\gamma}\, \partial_\gamma\xi^{3]\alpha}
\Big)\right.\nonumber\\&&
\left. -\frac{1}{2}\Big(\partial_\mu\xi^{1\rho}\, \partial_\rho\, \partial_\alpha \xi^{[2\beta}\partial_\beta\xi^{3]\alpha}+{\rm cycl. \ perm.}\Big) \right]\nonumber\\
&=& D\Big[
-\frac{1}{12}\partial_\mu\Big( {K^{[1}_\alpha}^\beta  \, {K^2_\beta}^\gamma\, {K^{3]}_\gamma}^\alpha \Big)
-\frac{1}{4}\, \partial_\mu \xi^{[1\nu}\, K^{2\rho\sigma}\ \partial_\nu\ K^{3]}_{\rho\sigma}
\Big]\nonumber\\ \nonumber\\
{[} A(\tilde\xi^1),[A(\tilde\xi^2),A(\tilde\xi^3)]]+{\rm cycl.\ perm.}
&=&0\nonumber\\ \nonumber\\
{[} A(\tilde\xi^1),[A(\tilde\xi^2),D(\xi^3)]]+{\rm cycl.\ perm.}&=&
D\left[ \frac{1}{2}\, \partial_\mu\Big(
\partial_\alpha\tilde\xi^{[1\beta}\,\partial_\beta\tilde\xi^{2]\gamma}\, \partial_\gamma\xi^{3\alpha}\Big) \right.\nonumber\\
&&
\left. -\frac{1}{2}\, \partial_\alpha\tilde\xi^{[1\beta}\ \partial_\gamma\ \partial_\beta\tilde\xi^{2]\alpha}\, \partial_\mu\xi^{3\gamma}- \partial_\alpha\tilde\xi^{[1\beta}\ \partial_\mu\ \partial_\beta\tilde\xi^{2]\gamma}\, \partial_\gamma\xi^{3\alpha} \right]
\nonumber\\
&=&
D\left[\frac{1}{12}\partial_\mu\Big( {{{\tilde K^{[1}}_\alpha}}{}^\beta  \, {\tilde K^2_\beta}{}^\gamma\, {K^{3]}_\gamma}^\alpha \Big)\right. \nonumber\\
&&
\left. -\frac{1}{4}\, {\tilde K^{[1}_\alpha}{}^\beta\ 
\partial_\gamma\  {\tilde K^{2]}_\beta}{}^\alpha\ \partial_\mu\xi^{3\gamma}-{\tilde K^{[1}_\alpha}{}^\beta \ \partial_\mu\  {\tilde K^{2]}_\beta}{}^\gamma\, \partial_{\gamma}\xi^{3\alpha}
\right]
\nonumber\\ \nonumber\\
{[}D(\xi^1),[D(\xi^2),A(\tilde\xi^3)]]+{\rm cycl.\ perm.}&=&
A\left[\frac{1}{4}\partial_\mu\ \Big(\xi^{[1\alpha}\, \xi^{2]}_\nu\, \partial_\alpha\tilde\xi^{3\nu} +\partial_\alpha\xi^{[1}_\nu\, \xi^{2]\alpha}\, \tilde\xi^{3\nu} 
\Big)\right. \nonumber\\
&&-\frac{1}{4}\, \partial_\mu\, \Big(\xi^{1\nu}\, \partial_\nu\, \partial_\alpha\xi^{[2\beta}\, \partial_\beta \tilde\xi^{3]\alpha}+{\rm{cycl.\ perm.}}\Big)\nonumber\\
&&\left. -\frac{1}{2}\, \Big( 2\, \partial_\alpha\xi^{[1\beta}\ \partial_\mu\ \partial_\beta\xi^{2]\gamma}\, \partial_\gamma \tilde\xi^{3\alpha}-\partial_\nu\, \partial_\alpha\xi^{[1\beta}\, \partial_\beta\, \xi^{2]\alpha}\ \partial_\mu\tilde\xi^{3\nu}\Big)
\right]
\nonumber\\
&=&
A\left[\frac{1}{4}\partial_\mu\ \Big(\xi^{[1\alpha}\, \xi^{2]}_\nu\, \partial_\alpha\tilde\xi^{3\nu} +\partial_\alpha\xi^{[1}_\nu\, \xi^{2]\alpha}\, \tilde\xi^{3\nu} -\frac{1}{2}\, 
\xi^{[1\nu}\, K^{2\rho\sigma}\ \partial_\nu\ \tilde K^{3]}_{\rho\sigma}
\Big)\right]\nonumber\\
&&\left. -\Big({K^{[1}_\alpha}^\beta  \ \partial_\mu \  {K^{2]}_\beta}^\gamma\, \partial_\gamma\tilde\xi^{3\alpha}
-\frac{1}{4}\, \partial_\nu \, {K^{[1}_\alpha}^\beta\, {K^{2]}_\beta}^\alpha\, \partial_\mu \tilde\xi^{3\nu}\Big) \right]\nonumber\\
\end{eqnarray}
We use the anti-symmetrization convention which assigns unit weight to each term in the permutation
sum. We also have the useful identities 
(under the assumption that contracted momenta
are neglected):
\begin{eqnarray}
{K^{1}_\alpha}^\beta  \, {K^2_\beta}^\gamma\, {K^{3}_\gamma}^\alpha&=&
\partial_\alpha\xi^{1\beta}\, \partial_\beta\xi^{2\gamma}\, \partial_\gamma\xi^{3\alpha}-\partial_\alpha\xi^{1\beta}\, \partial_\beta\xi^{3\gamma}\, \partial_\gamma\xi^{2\alpha}\nonumber\\
\xi^{[1\nu}\, K^{2\rho\sigma}\ \partial_\nu\ K^{3]}_{\rho\sigma}&=&-2\, \xi^{1\nu}\, \Big(
\partial^\rho\xi^{2\sigma}\ \partial_\nu\ \partial_\sigma\xi^3_\rho-\partial^\rho\xi^{3\sigma}\ \partial_\nu\ \partial_\sigma\xi^2_\rho
\Big)+{\rm{cycl.\ perm.}}\nonumber\\
{K^1_\alpha}^\beta\, {K^2_\beta}^\alpha&=&2\, \partial_\alpha \xi^{1\beta}\, \partial_\beta\xi^{2\alpha}
\, .
\end{eqnarray}

In the total derivative term,
 the Jacobiator of diffeomorphisms is anti-symmetric and linear in the three
parameters $\xi_{1,2,3}$. Moreover, the extra terms arise from a derivative with an external leg
acting on one of the three parameters. As such, it may allow for an embedding into a strongly
homotopy Lie algebra. This may be the case for the full algebra of diffeomorphisms and anti-symmetric
gauge transformations. We leave the closer study of the algebra, as well as the broader question to how this
algebraic structure is encoded in the tensor product of vertex operator algebras for future research.
(See e.g. \cite{Zeitlin:2009hc} for some inspiration.)


\end{document}